\documentclass[orivec]{llncs}
\usepackage[T1]{fontenc}
\usepackage{graphicx}
\graphicspath{{./fig/}}
\usepackage{hyperref}
\usepackage{url}
\usepackage{amsmath}
\usepackage{amssymb}

\let\doendproof\endproof \renewcommand\endproof{~\hfill$\qed$\doendproof}

\usepackage{color}

\newcommand{\out}{\mathit{output}}
\newcommand{\SA}{\mathsf{SA}}
\newcommand{\Occ}{\mathsf{Occ}}
\newcommand{\comp}{\mathcal{C}}
\newcommand{\ST}{\mathcal{T}}
\newcommand{\sdepth}{\mathsf{sdepth}}
\newcommand{\parent}{\mathsf{parent}}

\spnewtheorem{observation}{Observation}{\bfseries}{\itshape}

\begin{document}
\title{String Indexing for Maximal Gap-Bounded Occurrence Clusters}
\author{
  Waseem Akram\inst{1}\orcidID{0009-0007-4960-8673}
  \and
  Takuya Mieno\inst{2}\orcidID{0000-0003-2922-9434}
}
\authorrunning{W. Akram and T. Mieno}
\institute{
  Division of Medical Data Informatics, Human Genome Center, The Institute of Medical Science, The University of Tokyo, Japan\\
  \email{akram@hgc.jp}
  \and
  Graduate School of Informatics and Engineering, University of Electro-Communications, Japan\\
  \email{tmieno@uec.ac.jp}
}
\maketitle
\begin{abstract}
  Repeated occurrences of a pattern can form groups that reveal where repetition is concentrated in a text.
  We study how to find such groups efficiently
  when both the pattern $P$ and the distance threshold $G$ between occurrences are specified as query parameters.
  A maximal $G$-cluster of $P$ is a maximal block of at least two occurrences of $P$ whose consecutive starting positions differ by at most $G$.
  In this paper, we present indexes for reporting all cluster intervals and counting the clusters.
  For a text of length $n$ and a pattern of length $m$, we propose
  an $O(n\log n)$-space index that reports all maximal cluster intervals in $O(m+\log\log n+\out)$ time,
  where $\out$ is the number of reported intervals,
  and that counts the clusters in $O(m+\log\log n)$ time.
  Restricting occurrences to those fully contained in a query substring preserves the reporting time using $O(n\log^2 n)$ space.
  Our indexes combine suffix-tree decompositions with geometric reporting queries on consecutive gaps.
\end{abstract}

\section{Introduction}\label{sec:introduction}

In string indexing, a text $T$ is preprocessed to support efficient queries.
For example, given a pattern $P$, we can count its occurrences in $T$ or report their starting positions.
Let $n=|T|$, $m=|P|$, and let $\mathit{occ}$ be the number of occurrences of $P$
in $T$.
Classical suffix-tree indexes use $O(n)$ words and support 
reporting in $O(m+\mathit{occ})$ time~\cite{Weiner73}.
In some applications, it is useful to know both how often $P$ occurs in $T$ and how its occurrences are distributed.
These occurrences may be concentrated in a few regions or spread throughout the text.
In regulatory DNA, homotypic clusters group nearby binding sites
for the same transcription factor~\cite{EzerZA14}.
In natural-language corpora, successive occurrences of a word can
exhibit bursts and lulls~\cite{AltmannPM09}.
Motivated by local recurrence, we study a simple formulation based
on exact pattern occurrences and the distances between them.

For a nonnegative integer $G$, a \emph{maximal $G$-cluster} of $P$
is a maximal block of at least two occurrences in text order whose
consecutive starting positions differ by at most $G$.
The threshold bounds each consecutive gap; the cluster itself may
span a much longer interval.
If its first and last starting positions are $\ell$ and $r$, its
\emph{cluster interval} is $[\ell,r+m-1]$.
Both $P$ and $G$ are given as query parameters.
We study reporting all cluster intervals and counting the clusters.
We also study \emph{internal queries}, which report clusters within a query substring
$T[a..b]$.
In this version, only occurrences fully contained in $T[a..b]$ are
considered, and maximality is defined with respect to these
restricted occurrences.

A na\"ive solution reports all $\mathit{occ}$ occurrences in text order and
scans their consecutive gaps.
Using the online sorted range reporting structure~\cite{BrodalFGL09} on the suffix array, this takes
$O(m+\mathit{occ})$ time and $O(n)$ words.
However, the number of cluster intervals can be much smaller than
the number of occurrences.
A cluster containing $q$ occurrences has $q-1$ consecutive pairs
with gaps at most $G$, but contributes only one interval.
Thus,
enumerating either all occurrences or all consecutive occurrence pairs whose starting-position gaps are at most $G$
can require $\Theta(\mathit{occ})$ time even when the answer consists of one interval.
Our goal is to report cluster intervals in time depending on their
number, and to count the clusters without enumerating all occurrences.

In this paper, we work in the word RAM model with word size $\Theta(\log n)$.
We propose an $O(n\log n)$-word data structure
that reports all maximal cluster intervals in $O(m+\log\log n+\out)$ time and
that counts the clusters in $O(m+\log\log n)$ time~(Theorems~\ref{thm:maximal-reporting} and~\ref{thm:counting}),
where $\out$ denotes the number of reported cluster intervals.
For internal queries, we show that an $O(n\log^2 n)$-word index achieves the same reporting time~(Theorem~\ref{thm:substring-heavy-path}).
The key observation is that
an occurrence is an endpoint of a cluster if and only if it has a neighboring
occurrence within distance $G$ on one side but not the other.
By tracking each occurrence together with its predecessor and successor
along the heavy paths of the suffix tree, we represent these endpoint
conditions using $O(n\log n)$ \emph{weighted rectangles}.
This representation supports reporting cluster endpoints and counting left endpoints.

\paragraph*{Related Work.}
Navarro and Thankachan~\cite{NavarroT16} report consecutive occurrence
pairs with starting-position distances in a query-specified interval $[\alpha, \beta]$
using $O(n\log n)$ space and $O(m+k)$ query time,
where $k$ is the number of reported pairs.
For $\alpha=0$, Radoszewski~\cite{Radoszewski23} achieves the same query time in linear space.
Bille et al.~\cite{BilleGPRS22} report the $k$ closest consecutive
occurrence pairs in distance order
{using $O(n\log n)$ space and $O(m+k)$ query time.}
Akram and Mieno~\cite{AkramM25} extended both problems to occurrences
fully contained in a query substring, using $O(n\log^2 n)$ words
for top-$k$ queries and $O(n\log^{2+\epsilon}n)$ words for
gap-bounded queries, with $O(m+\log\log n+k)$ query time
for any fixed $\epsilon>0$.
Recently, Bals et al.~\cite{BalsCLPV26} proposed an $O(n)$-word index with $O(m)$ query time for counting consecutive occurrence pairs
whose distances lie in an interval $[\alpha, \beta]$ fixed during preprocessing.
Consecutive occurrence queries were also studied for two distinct
patterns~\cite{BilleGPS23} and on compressed texts~\cite{GawrychowskiGSS23}.
For $G=m$, the cluster intervals are exactly the inclusion-maximal
intervals longer than $m$ whose substrings are \emph{covered} by $P$,
i.e., every position in each interval lies within an occurrence
of $P$ fully contained in that interval.
Brodal and Pedersen~\cite{BrodalP00} use maximal blocks of
occurrence positions with consecutive gaps of length at most $m$
in their algorithm for maximal quasiperiodicities.
For $\alpha>1$, two occurrences of $P$ at starting-position distance $d$ form an \emph{$\alpha$-gapped repeat}
if and only if $m < d \le \alpha m$~\cite{KolpakovPPK17}.
Setting $G=\lfloor\alpha m\rfloor$ therefore places both
occurrences of every such repeat in the same maximal $G$-cluster.

\section{Preliminaries}\label{sec:preliminaries}

Let $\Sigma$ be an ordered alphabet.
For a string $X$, its length is denoted by $|X|$.
We denote by $X[i]$ the $i$th character and by $X[i..j]$
the substring from $i$ to $j$ for $1\le i \le j \le |X|$.
Throughout the paper, we fix a text $T$ of length $n$
and assume that $T[n]=\$$ is a sentinel that occurs exactly once in $T$.
Let $P$ be a nonempty query pattern of length $m$.
For a nonempty string $X$, let
$\Occ_T(X)=\{i : 1\le i\le n-|X|+1,\ T[i..i+|X|-1]=X\}$
be the set of starting positions of its occurrences in $T$.
We omit the subscript $T$ when it is clear from the context.
When we regard $\Occ_T(X)$ as a sequence, its elements are listed in increasing order.
Let $G$ be a nonnegative integer.
A \emph{$G$-cluster} of $P$ is a sequence
$C=(p_1,\ldots,p_k)$ of positions in $\Occ(P)$ such that
$k\ge2$, $p_1<\cdots<p_k$, and $p_{i+1}-p_i\le G$ for
every $1\le i<k$.
It is \emph{maximal} if it is not a proper subsequence of
any other $G$-cluster of $P$.
If $\ell$ and $r$ are its first and last starting positions,
its \emph{cluster interval} is $[\ell,r+m-1]$.
For a finite set $S=\{x_1,\ldots,x_s\}$ of positions, with
$x_1<\cdots<x_s$, a \emph{$G$-component} of $S$ is a set
$\{x_i,\ldots,x_j\}$, where $1\le i\le j\le s$, that is
maximal subject to
$x_{k+1}-x_k\le G$ for every $i\le k<j$.
We denote the set of $G$-components of $S$ by $\comp_G(S)$.
A component may consist of a single position.
Thus, the maximal $G$-clusters of $P$ correspond exactly to
the non-singleton $G$-components of $\Occ(P)$.
We write $c_G(S)=|\{C\in\comp_G(S):|C|\ge2\}|$ for the number
of non-singleton $G$-components of $S$.

\begin{definition}[Maximal gap-bounded cluster queries]
  Given a pattern $P$ and a nonnegative integer $G \le n$,
  a \emph{maximal gap-bounded cluster query (MGC query)}
  reports the cluster intervals of all maximal $G$-clusters of $P$.
  A counting query takes the same input and returns $c_G(\Occ(P))$.
\end{definition}

We use the word RAM model with word size $w=\Theta(\log n)$
and measure space in words.
Characters and query integers are assumed to fit in one word.
All asymptotic bounds are stated for sufficiently large size
parameters; empty and constant-size instances are handled directly.

The suffix tree $\ST$ of $T$ is the compact trie of the suffixes of $T$~\cite{Weiner73}.
Since the sentinel $\$$ occurs only at the end of $T$,
each suffix corresponds to a distinct leaf.
We label this leaf with the starting position of the suffix.
For a node $v$, let $\Occ_v$ be the set of leaf labels
in the subtree rooted at $v$.
The string depth $\sdepth(v)$ is the length of the string obtained
by concatenating the edge labels on the path from the root to $v$.
For a non-root node $v$, its parent is denoted by $\parent(v)$.
For convenience, for the root $r$, we define $\parent(r) = \bot$ and $\sdepth(\bot) = -\infty$.
Let $\SA[1..n]$ denote the suffix array of $T$, which lists the
starting positions of its suffixes in lexicographic order.
The \emph{suffix-array interval} of $v$ is the contiguous range
of indices in the suffix array $\SA$ whose entries belong to $\Occ_v$.
At each node $v$, we store this interval.
For a pattern $P$ occurring in $T$, its \emph{locus} is a node
or point on an edge reached by reading $P$ from the root.
If the locus is a node, we call this node the
\emph{representative node} of $P$.
Otherwise, the representative node is the child endpoint
of the edge containing the locus.
For the representative node $v$, we have $\Occ_v=\Occ(P)$.
We use an $O(n)$-word representation of $\ST$ with constant-time
static dictionaries (e.g., minimal perfect hash) for selecting outgoing edges.
Thus, in $O(|P|)$ time, we can find the representative node of $P$
or determine that $P$ does not occur in $T$.

For each internal node of $\ST$, choose a child with the largest
number of descendant leaves as its \emph{heavy child}, breaking ties arbitrarily.
The edge to the heavy child is called a \emph{heavy edge},
and all other edges are called \emph{light edges}.
The heavy edges partition the nodes into \emph{heavy paths};
each path is extended as far as possible and may consist of
a single node.
Let $H=(v_0,\ldots,v_{h_H})$ be a heavy path, listed from its
head $v_0$ to its final leaf $v_{h_H}$.
Let $\lambda_H(v_i)=i$ be the node index of $v_i$ and let $s_H = |\Occ_{v_0}|$.
The occurrence sets are nested along the path:
$\Occ_{v_0}\supseteq\Occ_{v_1}\supseteq\cdots \supseteq\Occ_{v_{h_H}}$.
Any root-to-leaf path contains at most $\lfloor\log n\rfloor$ light edges since
every light child has at most half as many descendant leaves as its parent.
Each position belongs to the occurrence sets of
at most $1+\lfloor\log n\rfloor$ heavy-path heads.
Counting these memberships over all $n$ positions gives
\begin{equation}\label{eq:headsum}
  \sum_{H\text{: heavy path}} s_H=O(n\log n).
\end{equation}

\section{Tools}
\label{sec:shared}

This section gives algorithmic tools and combinatorial observations used in the proposed method.
For a set $S$ of integer keys and an integer $x$, a predecessor query
returns the largest key at most $x$, and a successor query returns
the smallest key at least $x$.
If no such key exists, the query returns $\mathit{nil}$.
\begin{lemma}[\cite{cpm/0001G15}, Proposition 2]\label{lem:pred}
  Given a set $S \subseteq \{0, 1, \ldots, n\}$ of \emph{sorted} integer keys,
  we can construct, in $O(|S|)$ time, a data structure of size $O(|S|)$
  that supports predecessor and successor queries
  in $O(\log\log n)$ time.
\end{lemma}

Let $N=n^{O(1)}$ horizontal segments have integer coordinates
in a universe of size $U=n^{O(1)}$ and pairwise distinct heights.
A \emph{sorted segment intersection} query is,
given a closed vertical segment as a query,
to report the horizontal segments intersecting the query segment
in increasing height order.
\begin{lemma}
  \label{lem:sorted-segments}
  There is a data structure of size $O(N)$
  that answers any sorted segment intersection query
  in $O(\log\log n+z)$ time, where $z$ is the number of outputs.
  The data structure can be constructed in $O(N\log N)$ time.
\end{lemma}
\begin{proof}
  We use Chan's sorted \emph{vertical-ray} intersection structure~\cite[Corollary~4.2(a)]{Chan13}.
  For a query segment $\{x\}\times[a,b]$, report the horizontal
  segments intersecting the upward ray from $(x,a)$ in increasing height order.
  Stop when a reported segment has height greater than $b$, or when no segment remains.
  At most $z+1$ segments are examined, giving $O(\log\log n+z)$ query time.
\end{proof}

For a nonempty position set $S=\{p_1,\ldots,p_s\} \subseteq [1, n]$, with
$p_1<\cdots<p_s$, let
$d_i=p_{i+1}-p_i$ for $1\le i<s$ and set $d_0=d_s=n+1$.
For notation, write $p_0=-\infty$ and $p_{s+1}=+\infty$.
Let $\mathbf{1}$ denote the indicator function,
where $\mathbf{1}[Q]$ is one if $Q$ holds and zero otherwise.

\begin{lemma}\label{lem:endpoint-tests}
  \label{lem:count-left-endpoints}
  Let $G$ be an integer with $0 \le G \le n$.
  A position $p_i$ is an endpoint of a non-singleton $G$-component
  of $S$ if and only if
  $\min\{d_{i-1},d_i\}\le G<\max\{d_{i-1},d_i\}$.
  It is a left endpoint if and only if $d_i\le G<d_{i-1}$.
  Namely, $c_G(S)=\sum_{i=1}^{|S|}\mathbf{1}[d_i\le G<d_{i-1}]$.
\end{lemma}

\begin{proof}
  A position is an endpoint of a non-singleton component exactly
  when it belongs to the same component as exactly one of its
  two neighbors.
  This means that exactly one of $d_{i-1}$ and $d_i$ is at most $G$,
  which is equivalent to
  $\min\{d_{i-1},d_i\}\le G<\max\{d_{i-1},d_i\}$.
  It is a left endpoint if and only if the gap to its successor
  is at most $G$ and the gap to its predecessor is greater than $G$.
  In particular, positions whose $G$-components are singletons are excluded, including when
  $s=1$.
  Every non-singleton component has exactly one left endpoint,
  which proves
  $c_G(S)=$
  $\sum_{i=1}^{|S|}\mathbf{1}[d_i\le G<d_{i-1}]$.
\end{proof}

At each node $v$
of the suffix tree, associate each occurrence
$p\in\Occ_v$ with the triple $(p^-,p,p^+)$;
here, $p^-$ is the occurrence immediately preceding $p$ in
$\Occ_v$, or $-\infty$ if $p$ is the first occurrence;
similarly, $p^+$ is the occurrence immediately following $p$
in $\Occ_v$, or $+\infty$ if $p$ is the last occurrence.
As we move down a heavy path, occurrences may disappear,
and the predecessor or successor of a remaining occurrence may change.
For a fixed triple $(p^-,p,p^+)$, a \emph{lifetime} is a maximal
interval of node indices of the heavy path
such that, at every node in the interval,
$p$ is present and its neighbors are $p^-$ and $p^+$.
A \emph{record} stores the triple together with its lifetime.
The heavy-path representations in~\cite{NavarroT16,AkramM25} use consecutive occurrence \emph{pairs}.
Here, we use the \emph{triples} defined above and obtain the following bounds:
\begin{lemma}\label{lem:local-lifetimes}
  For each heavy path $H$, the number of triple records is $O(s_H)$.
  There are $O(n\log n)$ records over all heavy paths.
  They can be computed in $O(n\log^2 n)$ time using $O(n\log n)$ words.
\end{lemma}
\begin{proof}
  At the head $v_0$ of $H$, there are $s_H = |\Occ_{v_0}|$ triples.
  Deleting an occurrence removes its triple and changes only the
  triples centered at its immediate predecessor and successor.
  Since each occurrence is deleted at most once along $H$ from the top to the bottom,
  there are $O(s_H)$ records.
  Once a triple disappears, it cannot reappear, so its lifetime
  is a single interval of node indices of $H$.
  For construction, sort the occurrence list $\Occ_{v_0}$ for each heavy path
  and maintain it as a doubly linked list.
  At each transition, enumerate the leaves of the light-child
  subtrees and delete their positions.
  Locate each position by binary search in the sorted list;
  the remaining updates take constant time per deletion.
  Record the corresponding node indices, discarding records
  created and destroyed within the same transition.
  This takes $O(s_H\log n)$ time and $O(s_H)$ words per path.
  Summing over all paths and applying Equation~\eqref{eq:headsum}
  gives the stated bounds.
\end{proof}

\begin{figure}[tb]
  \centering
  \includegraphics[width=\linewidth]{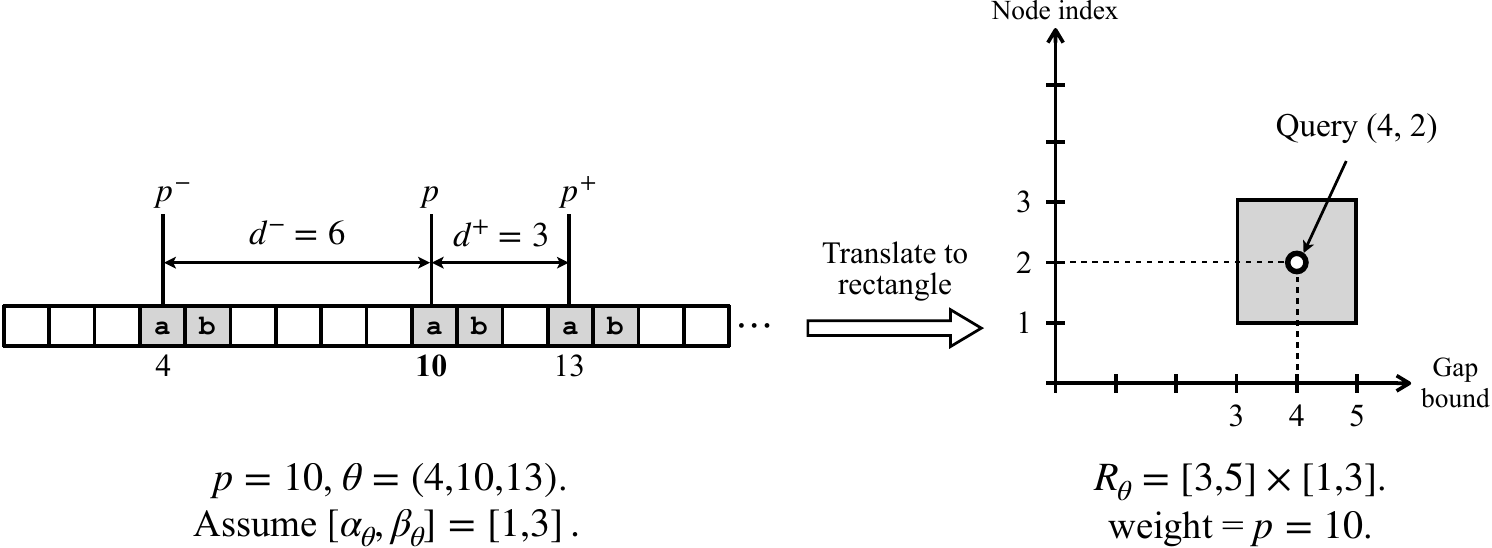}
  \caption{Illustration for the rectangle translation.
    Consider the occurrences $4, 10, 13,\ldots$ of $\mathtt{ab}$ and let $p = 10$.
    Then the triple centered at $p$ is $\theta = (4, 10, 13)$.
    Assume that the triple $\theta$ has lifetime $[1, 3]$.
    Since $(d^{\min}_\theta,d^{\max}_\theta) = (3, 6)$, the corresponding rectangle is
    $R_\theta = [3, 5]\times[1, 3]$ with weight $p = 10$. 
  }
  \label{fig:rect}
\end{figure}

\paragraph*{Representing endpoints by rectangles.}
For a record $\theta=(p^-,p,p^+)$ on $H$, let $[\alpha_\theta,\beta_\theta]$ be its lifetime.
Let $d^-_\theta$ and $d^+_\theta$ be its left and right gaps, using $n+1$ for a missing neighbor.
Define $d^{\min}_\theta=\min\{d^-_\theta,d^+_\theta\}$ and $d^{\max}_\theta=\max\{d^-_\theta,d^+_\theta\}$.
If $d^{\min}_\theta<d^{\max}_\theta$, create the rectangle
$R_\theta= [d^{\min}_\theta,d^{\max}_\theta-1] \times[\alpha_\theta,\beta_\theta]$ with weight $p$.
Otherwise, create no rectangle (see Fig.~\ref{fig:rect}).
Let $\mathcal{R}_H$ be the resulting family of rectangles, and define its subfamily
$\mathcal{L}_H= \{R_\theta\in\mathcal{R}_H : d^+_\theta<d^-_\theta\}$.
Records are kept distinct even if their geometric rectangles coincide.

Fix a node $v\in H$.
For each position $p\in\Occ_v$, exactly one triple centered at $p$
has a lifetime containing $v$'s index $\lambda_H(v)$ in $H$.
By Lemma~\ref{lem:endpoint-tests}, the weights of the rectangles
in $\mathcal{R}_H$ containing point $(G,\lambda_H(v))$ are exactly
the endpoints of the non-singleton components of $\Occ_v$.
Similarly, the rectangles in $\mathcal{L}_H$ containing this point
correspond exactly to their left endpoints.

Let $N_H=|\mathcal{R}_H|$.
Lemma~\ref{lem:local-lifetimes} gives $N_H=O(s_H)$ and $\sum_H N_H=O(n\log n)$.
The non-singleton components form disjoint blocks in the sorted
occurrence list.
Thus, if all their endpoints are reported in increasing position
order, they alternate between left and right endpoints.
Pairing consecutive endpoints $\ell,r$ and reporting
$[\ell,r+m-1]$ gives exactly the maximal $G$-cluster intervals.

\paragraph*{Structures for short patterns.}
We use the following fact for short patterns.

\begin{lemma}\label{lem:shallow-nodes}
  For an integer $\tau\ge1$, let
  $\mathcal{V}_\tau= \{v : \sdepth(\parent(v))<\tau\}$.
  The representative node of every occurring pattern of length
  $m<\tau$ belongs to $\mathcal{V}_\tau$, and
  $\sum_{v\in\mathcal{V}_\tau}|\Occ_v| = O(n\tau)$.
\end{lemma}
\begin{proof}[sketch]
  Each root-to-leaf path contains at most $\tau$ non-root nodes
  in $\mathcal V_\tau$, since their parent string depths are
  distinct integers in $\{0,\ldots,\tau-1\}$.
\end{proof}

Storing an $O(|\Occ_v|)$-word structure at each
node $v\in\mathcal{V}_\tau$ uses $O(n\tau)$ words in total.
The lists $\Occ_v$ for $v\in \mathcal V_\tau$
can be enumerated in
$O(n\tau)$ time and sorted in $O(n\tau\log n)$ time.
In each application below, we choose $\tau$ so that the query algorithm
using the rectangle structures built for the heavy paths has an additive
overhead of $O(\tau)$.
For $m \ge \tau$, this overhead is absorbed into $O(m)$.
For $m < \tau$, we use the structure stored at the representative node.
The structures stored at nodes in $\mathcal V_\tau$ and the choice of $\tau$
are specified
in the respective applications.

\section{Maximal gap-bounded occurrence clusters}
\label{sec:mxml-gap-bounded}

This section gives a proof of our main theorem:
\begin{theorem}\label{thm:maximal-reporting}
  There is an $O(n\log n)$-word index that answers MGC queries in
  $O(m+\log\log n+\out)$ time, where $\out$ is the number of reported cluster intervals.
\end{theorem}

If $P$ does not occur in $T$, a reporting query returns no intervals
and a counting query returns zero.
Otherwise, let $v$ be the representative node of $P$.

We first describe an algorithm with $O(m+|\Occ(P)|)$ query time.
We build the linear-space online sorted range reporting structure
of Brodal et al.~\cite{BrodalFGL09} on the suffix array $\SA$.
After locating $v$, we use its suffix-array interval to report
the occurrences of $P$ in increasing text-position order.
We then scan these positions, starting a new component whenever the
gap between consecutive positions exceeds $G$.
For each component containing at least two positions, we report
its cluster interval.
The total query time is $O(m+|\Occ(P)|)$.
When $m\ge |\Occ(P)|$, the query time is $O(m)$.
We next describe indexes that report the cluster endpoints directly.

\subsection{A reporting structure at each suffix-tree node}
\label{ssec:node-endpoints}

\begin{lemma}\label{lem:node-endpoints}
  For any node $v$ of the suffix tree,
  an $O(|\Occ_v|)$-word data structure supports the following query:
  Given an integer $G$ with $0\le G\le n$ and an interval $[\ell,r]$,
  it reports, in increasing text-position order,
  all positions in $[\ell,r]$ that are endpoints of
  non-singleton $G$-components of $\Occ_v$.
  The structure supports the query in $O(\log\log n+z)$ time,
  where $z$ is the number of outputs.
  The structure also supports predecessor and successor queries on $\Occ_v$ in $O(\log\log n)$ time.
\end{lemma}

\begin{proof}
  Let $\Occ_v=\{p_1,\ldots,p_{|\Occ_v|}\}$ with $p_1<\cdots<p_{|\Occ_v|}$,
  and use the gap convention from Section~\ref{sec:shared} (e.g., $d_i=p_{i+1}-p_i$).
  For each $p_i$, we define
  $d_i^{\min}=\min\{d_{i-1},d_i\}$ and 
  $d_i^{\max}=\max\{d_{i-1},d_i\}$.
  If $d_i^{\min}<d_i^{\max}$, create the horizontal segment
  $[d_i^{\min},d_i^{\max}-1]\times\{p_i\}$.
  Otherwise, create no segment.
  By Lemma~\ref{lem:endpoint-tests}, the vertical query segment
  $\{G\}\times[\ell,r]$ intersects exactly the segments corresponding
  to the desired endpoints.
  The segments have distinct heights,
  so they can be reported in increasing text-position order
  in $O(\log\log n+z)$ time by Lemma~\ref{lem:sorted-segments}.
  In addition, store all positions of $\Occ_v$ as keys in the
  structure of Lemma~\ref{lem:pred}.
  This supports the predecessor/successor queries.
  There are at most $|\Occ_v|$ segments and exactly $|\Occ_v|$ position keys,
  so the total space is $O(|\Occ_v|)$ words.
\end{proof}

Building the structure of Lemma~\ref{lem:node-endpoints} at every suffix-tree node
uses $O(\sum_v |\Occ_v|)=O(n^2)$ words.
To answer an MGC query, query the structure at $v$ with
interval $[1,n]$.
Pair the reported endpoints as described in
Section~\ref{sec:shared}.
Since there are exactly $2\out$ endpoints, this gives
$O(m+\log\log n+\out)$ query time.

\subsection{An $O(n\log n)$-word index}

We use a 2-d rectangle stabbing data structure of~\cite{ChanNRT22}.
\begin{lemma}[\cite{ChanNRT22}]\label{lem:rect_stab}
  Let $\mathcal{R}$ be a set of $N$ axis-aligned half-open rectangles in 2-d rank space, with distinct priorities in $\{1,\ldots,N\}$.
  There is a linear-space data structure that, given a query point $q$, reports the rectangles in $\mathcal{R}$ containing $q$ in decreasing order of priority in $O(\log_w N+z)$ time where $w$ is the word size and $z$ is the number of rectangles to report.
\end{lemma}
Now we prove Theorem~\ref{thm:maximal-reporting}.
\begin{proof}[of Theorem~\ref{thm:maximal-reporting}]
  For each heavy path $H$, consider the family $\mathcal{R}_H$
  of endpoint rectangles defined in Section~\ref{sec:shared}.
  Recall that $N_H=|\mathcal{R}_H|=O(s_H)$.
  At a node $v\in H$, the weights of the rectangles containing
  point $(G,\lambda_H(v))$ are exactly the endpoints of the non-singleton $G$-components of $\Occ_v$.

  We use the linear-space 2-d rectangle
  stabbing structure of Lemma~\ref{lem:rect_stab}.
  For a rectangle of weight $p$, assign priority $n+1-p$.
  Break ties by record identifiers and replace these ordered
  priorities by their ranks.
  Decreasing priority order then corresponds to increasing
  text-position order.

  We replace each closed coordinate interval $[a,b]$ by $[a,b+1)$
  and rank the distinct endpoints in each coordinate.
  Query coordinates are mapped to predecessor ranks in the
  sorted arrays of distinct endpoints $E_x$ and $E_y$,
  with rank zero if no predecessor exists.
  This preserves containment of integer query points and matches the half-open convention of Lemma~\ref{lem:rect_stab}.
  The resulting coordinate universe has size $O(N_H)$.
  The arrays $E_x$ and $E_y$ use $O(N_H)$ words, and query coordinates are converted by binary search in $O(\log n)$ total time.

  Build the stabbing structure of Lemma~\ref{lem:rect_stab}
  on these transformed rectangles with their ranked priorities.
  Its online reporting procedure does not require the output size in advance and reports rectangles in decreasing priority order.
  After converting query point $(G,\lambda_H(v))$, we continue this procedure until all containing rectangles have been reported.

  For each $p\in\Occ_v$, exactly one triple record on $H$ has $p$ as its middle entry and a lifetime containing the node index $\lambda_H(v)$.
  Thus, the reported weights are distinct and give precisely the $2\out$ endpoints in increasing text-position order.
  Pairing consecutive endpoints yields the cluster intervals.
  After coordinate conversion, the geometric query takes $O(\log_w n+\out)$ time.
  Including coordinate conversion, the time is $O(\log n+\out)$.
  Since $N_H=O(s_H)$ and $s_H\ge1$, the structures for $H$ use $O(s_H)$ words, including constant overhead.
  By Equation~\eqref{eq:headsum}, their total space is $O(\sum_H s_H)=O(n\log n)$ words.
  Including the pattern search, the query time is therefore $O(m+\log n+\out)$.

  For short patterns,
  set $\tau=\lceil\log n\rceil$ and build the structure of
  Lemma~\ref{lem:node-endpoints} at every node in $\mathcal{V}_\tau$.
  By Lemma~\ref{lem:shallow-nodes}, these structures use
  $O(n\tau)=O(n\log n)$ words in total.
  If $m<\tau$, the representative node belongs to $\mathcal{V}_\tau$.
  We query the structure of Lemma~\ref{lem:node-endpoints} at $v$ with the position interval $[1,n]$
  and obtain $O(m+\log\log n+\out)$ query time.
  If $m \ge \tau$, we query the rectangle stabbing structure
  for the heavy path containing $v$.
  Its $O(\log n)$ overhead is absorbed by $O(m)$, giving
  $O(m+\out)$ query time.
  This proves Theorem~\ref{thm:maximal-reporting}.
\end{proof}

Note that we do not state a construction-time bound for the index in Theorem~\ref{thm:maximal-reporting},
as Chan et al.~\cite{ChanNRT22} do not specify the construction time of their rectangle stabbing structure.

\subsection{Counting maximal clusters}\label{ssec:counting}

We now count the maximal clusters in the full occurrence set
$\Occ(P)$ without enumerating their intervals.

\begin{theorem}\label{thm:counting}
  There is a data structure of size $O(n\log n)$ that returns $c_G(\Occ(P))$ in $O(m+\log\log n)$ time.
  The data structure can be constructed in $O(n\log^2 n)$ time.
\end{theorem}
\begin{proof}

  By Lemma~\ref{lem:endpoint-tests}, the required count at a node
  $v\in H$ is the number of rectangles in $\mathcal{L}_H$ containing
  $(G,\lambda_H(v))$.
  We use the following geometric data structure.

  \begin{lemma}
    \label{lem:count-stabbing}
    For $N=O(n)$ closed rectangles
    with integer coordinates in $\{0,\ldots,n\}$,
    there is a data structure of size $O(N)$ that counts the rectangles containing an integer query point
    in $O(\log N)$ time.
    The data structure can be constructed in $O(N\log N)$ time.
  \end{lemma}

  \begin{proof}[of Lemma~\ref{lem:count-stabbing}]
    For an integer query point $(x,y)$, a rectangle
    $[x^-,x^+]\times[y^-,y^+]$ contributes
    $(\mathbf{1}[x^-\le x]
    -\mathbf{1}[x^++1\le x])
    (\mathbf{1}[y^-\le y]
    -\mathbf{1}[y^++1\le y])$
    to the stabbing count.
    By expanding this,
    the contribution of $[x^-,x^+]\times[y^-,y^+]$ is written as
    \[
      \begin{aligned}
&\mathbf{1}\bigl[x^{-}\le x \land y^{-}\le y\bigr]
-\mathbf{1}\bigl[x^{+}+1\le x \land y^{-}\le y\bigr] \\
&\quad
-\mathbf{1}\bigl[x^{-}\le x \land y^{+}+1\le y\bigr]
+\mathbf{1}\bigl[x^{+}+1\le x \land y^{+}+1\le y\bigr].
      \end{aligned}
    \]

    We maintain four \emph{multisets} $S_1,S_2,S_3,S_4$.
    For each rectangle $[x^-,x^+]\times[y^-,y^+]$,
    we insert $(x^-,y^-)$, $(x^++1,y^-)$, $(x^-,y^++1)$, and $(x^++1,y^++1)$ into $S_1,S_2,S_3,S_4$, respectively.
    For each multiset, we sort its points by their first coordinates and write the resulting sequence as
    $(a_1,b_1),\ldots,(a_N,b_N)$, where $a_1\le\cdots\le a_N$.
    To count the rectangles containing a query point, we perform one two-dimensional range counting query on each of the four multisets.
    Specifically, we store the array $A[1..N]$, with $A[i]=a_i$, and the array $D$ of distinct second-coordinate values in increasing order.
    Let $B_\mathrm{rnk}[i]$ be the rank of $b_i$ in $D$.
    For an integer query point $(x,y)$, binary search in $A$ and $D$ determines
    $t=|\{i:A[i]\le x\}|$ and $r=|\{j:D[j]\le y\}|$.
    The number of points $(a_i,b_i)$ with $a_i\le x$ and $b_i\le y$ is the number of entries at most $r$ in $B_\mathrm{rnk}[1..t]$.
    If $t=0$ or $r=0$, the count is zero.
    Otherwise, these entries correspond exactly to the points $(i, B_\mathrm{rnk}[i])$ in the rectangle $[1,t]\times[1,r]$.
    It is known that a \emph{wavelet tree} can answer such a range counting query in $O(\log N)$ time~(e.g., \cite[Sect.~4.3]{Navarro12}).
    We therefore build a wavelet tree on $B_\mathrm{rnk}$ during preprocessing.

    Let $c_j$ be the count obtained from $S_j$.
    Then the number of rectangles containing $(x,y)$ is $c_1-c_2-c_3+c_4$.
    For each multiset, the coordinate arrays use $O(N)$ words, and the wavelet tree uses $O(N\log N)$ \emph{bits}.
    Thus, the four structures use $O(N)$ words in total.
    Sorting, rank conversion, and wavelet-tree construction take $O(N\log N)$ time in total.
    The total query time is $O(\log N)$.
  \end{proof}

  Apply Lemma~\ref{lem:count-stabbing} separately to each $\mathcal{L}_H$.
  Since $|\mathcal L_H|\le N_H=O(s_H)$ and $s_H\ge1$,
  the structure for $H$ uses $O(s_H)$ words and can be
  constructed in $O(s_H\log n)$ time.
  Equation~\eqref{eq:headsum} therefore gives $O(n\log n)$ total
  space and
  $O(\log n\sum_H s_H)=O(n\log^2 n)$ construction time.
  The rectangle records themselves can be constructed within the
  same bound by Lemma~\ref{lem:local-lifetimes}.

  A query first locates the representative node $v\in H$ and then
  counts the rectangles of $\mathcal{L}_H$ containing
  $(G,\lambda_H(v))$.
  The total query time is $O(m+\log n)$.

  For short patterns,
  set $\tau=\lceil\log n\rceil$.
  At each node $v\in\mathcal{V}_\tau$, write the occurrence positions
  as $p_1<\cdots<p_s$.
  Let $d_i=p_{i+1}-p_i$ for $1\le i<s$, and set $d_0=d_s=n+1$.
  Each non-singleton $G$-component has exactly one left endpoint,
  so it suffices to count these endpoints.
  An occurrence $p_i$ is such an endpoint if and only if $d_i\le G<d_{i-1}$:
  it must connect to the next occurrence but not to the previous one.
  Thus, each $i$ with $d_i<d_{i-1}$ contributes one to
  $c_G(\Occ_v)$ precisely for $G\in[d_i,d_{i-1})$.
  We compute these counts by sweeping the interval endpoints
  in increasing order.
  For each $i$ with $d_i<d_{i-1}$, create an event of value $+1$
  at threshold $d_i$ and an event of value $-1$
  at threshold $d_{i-1}$.
  These mark where the occurrence starts and stops contributing
  to the count, respectively.
  Discard events whose thresholds exceed $n$, since $G\le n$.
  Sort the events by their thresholds and compute the cumulative
  sum of their values, storing the sum after processing
  all events at each distinct threshold.
  The sum at the largest threshold at most $G$ is then
  $c_G(\Occ_v)$.
  Using Lemma~\ref{lem:pred}, we find this threshold and return
  its stored count in $O(\log\log n)$ time,
  or return zero if no predecessor exists.
  Each table has $O(|\Occ_v|)$ entries and can be built in
  $O(|\Occ_v|\log n)$ time, including sorting the occurrence
  list and events.
  By Lemma~\ref{lem:shallow-nodes}, the total space is
  $O(n\tau)=O(n\log n)$ words, and the total construction time is
  $O(n\tau\log n)=O(n\log^2 n)$.

  If $m<\tau$, we query
  the threshold table at the representative node
  in $O(m+\log\log n)$ total time.
  If $m \ge \tau$, we query the structure of Lemma~\ref{lem:count-stabbing} for the heavy path containing $v$.
  Its $O(\log n)$ overhead is absorbed by $O(m)$, giving
  $O(m)$ query time.
  Together with the construction bounds above, this proves
  Theorem~\ref{thm:counting}.
\end{proof}

The reporting and counting structures can be stored together
within $O(n\log n)$ words, supporting both types of queries
with their respective time bounds.

\section{Internal maximal gap-bounded occurrence clusters}
\label{sec:substring}

An \emph{internal MGC query} (IMGC query) takes a pattern $P$, a nonnegative
integer $G \le n$, and an interval $[a,b]$ with $1\le a\le b\le n$,
and reports the intervals of all $G$-clusters maximal among the
occurrences of $P$ fully contained in $T[a..b]$.
This section shows the following theorem.

\begin{theorem}\label{thm:substring-heavy-path}
  There is a data structure of size $O(n\log^2 n)$ that answers
  IMGC queries in $O(m+\log\log n+\out)$ time,
  where $\out$ is the number of reported intervals.
  The data structure can be constructed in $O(n\log^3 n)$ time.
\end{theorem}

For internal queries,
we need to report rectangles containing a query point in sorted weight order,
subject to a \emph{query-specified weight interval}.
Let $\mathcal R$ be a family of $N$ closed axis-parallel rectangles
with integer coordinates in $\{0,\ldots,n\}$ and integer weights.
Assume that $N$, all weights, and all query parameters fit in one word.
A \emph{weighted sorted rectangle stabbing query} is,
given point $(x, y)$ and weight interval $[\omega^-,\omega^+]$,
to report the rectangles containing point $(x, y)$ with weights in $[\omega^-,\omega^+]$
in nondecreasing order of weight.

\begin{lemma}
  \label{lem:weighted-sorted-stabbing}
  There is a data structure of size $O(N\log N)$ that answers
  weighted sorted rectangle stabbing queries in $O(\log N\log\log n+z)$ time,
  where $z$ is the number of outputs.
  The data structure can be constructed in $O(N\log^2 N)$ time.
\end{lemma}
\begin{proof}
  First, we assign distinct identifiers and rank the records
  lexicographically by weight and identifier.
  Let $e_1<\cdots<e_k$ be the distinct $x$-coordinates of rectangle endpoints.
  We build a segment tree whose leaves represent the singleton intervals $\{e_i\}$
  and the open intervals $(e_i,e_{i+1})$, in their order on the $x$-axis.
  We store each $x$-projection in its $O(\log N)$ canonical nodes.
  At each such node, represent a rectangle $[x^-,x^+]\times[y^-,y^+]$ of rank $\rho$
  by the horizontal segment $[y^-,y^+]\times\{\rho\}$ and apply Lemma~\ref{lem:sorted-segments}.

  Given a query, we first translate the weight interval $[\omega^-,\omega^+]$ into a rank interval $[\rho^-,\rho^+]$ by binary searches.
  We then visit the $O(\log N)$ nodes on the search path for $x$
  and query their structures with $\{y\}\times[\rho^-,\rho^+]$.
  For each rectangle containing $(x,y)$, exactly one node on the search path for $x$ stores its $x$-projection.
  Materializing the sorted outputs in arrays therefore takes $O(\log N\log\log n+z)$ total time.
  Finally, we merge the sorted arrays using a dynamic fusion tree~\cite{PatrascuT14} as follows:
  For each array with unreported entries, the fusion tree stores its first unreported rank together with a pointer to the array.
  We repeatedly extract the minimum rank and report the corresponding rectangle.
  If the same array has another entry, we insert its next rank.
  Since each array is sorted, this procedure reports all rectangles in increasing rank order.
  There are $O(\log N) = w^{O(1)}$ keys, so each update and minimum selection takes $O(1)$ time.
  The merge takes $O(\log N+z)$ time, including initialization.

  There are $O(N\log N)$ segments in total.
  Lemma~\ref{lem:sorted-segments} gives $O(N\log N)$ space
  and $O(N\log^2 N)$ construction time,
  since the structure at each segment-tree node contains at most $N$ segments.
  The output arrays use $O(N)$ additional words during a query.
\end{proof}

\begin{proof}[of Theorem~\ref{thm:substring-heavy-path}]
  Set $\ell=a$ and $r=b-m+1$; return no intervals if $\ell>r$
  or $P$ does not occur.
  Otherwise, let $v$ be the representative node of $P$ and let
  $\{p'_1, \ldots , p'_{s'}\} := \Occ_v\cap[\ell,r]$ with $p'_1<\cdots<p'_{s'}$.
  At most four predecessor and successor queries in $\Occ_v$
  find the first two and last two positions of
  $\Occ_v\cap[\ell,r]$, or determine that $s'<2$,
  in which case the answer is empty.
  For $1<i<s'$, the predecessor and successor of $p'_i$ are
  the same in $\Occ_v$ and $\Occ_v\cap[\ell,r]$.
  Let $E$ be the set of positions in $[p'_1+1,p'_{s'}-1]$
  that are endpoints of non-singleton $G$-components of
  $\Occ_v$, and let $z=|E|$.
  By Lemma~\ref{lem:endpoint-tests},
  the increasing sequence of endpoints of the non-singleton $G$-components of $\Occ_v\cap[\ell,r]$
  can be obtained by 
  (1) listing the elements of $E$ in increasing order;
  (2) prepending $p'_1$ if $p'_2-p'_1\le G$; and
  (3) appending $p'_{s'}$ if $p'_{s'}-p'_{s'-1}\le G$.
  Pair consecutive entries $x_1,x_2$ of this sequence and report $[x_1,x_2+m-1]$.
  Thus $z\le 2\out$.

  For each heavy path $H$, build the structure of Lemma~\ref{lem:weighted-sorted-stabbing} on $\mathcal R_H$
  (defined in Section~\ref{sec:shared}).
  Also build a wavelet tree~\cite{Navarro12} on the suffix array $\SA$;
  range predecessor and successor queries on the suffix-array interval of $v$ find the first two and last two positions of
  $\Occ_v\cap[\ell,r]$, or determine that $s'<2$, in $O(\log n)$ total time.

  Given query parameters $P$ and $G$, we first find the representative node $v$ of $P$.
  Let $H$ be the heavy path containing $v$.
  We then query $\mathcal R_H$ at point $(G,\lambda_H(v))$ with weight interval $[p'_1+1,p'_{s'}-1]$, skipping an empty interval.
  The reported rectangle weights are exactly the elements of $E$.
  The total query time is therefore $O(m+\log n\log\log n+\out)$.
  Since $N_H=O(s_H)$, Equation~\eqref{eq:headsum} gives
  $O(n\log^2 n)$ total space and $O(n\log^3 n)$ construction time.
  These bounds include rectangle generation by
  Lemma~\ref{lem:local-lifetimes} and the wavelet tree, which uses
  $O(n)$ words and is built in $O(n\log n)$ time.

  Finally, set $\tau=\lceil\log n\log\log n\rceil$ and store
  the structure of Lemma~\ref{lem:node-endpoints} at every node in $\mathcal{V}_\tau$.
  For $m<\tau$, use the predecessor and successor queries of
  the structure at $v$ to find the first two and last two positions
  of $\Occ_v\cap[\ell,r]$, or determine that $s'<2$.
  When $s'\ge2$, query the structure of Lemma~\ref{lem:node-endpoints} at $v$
  with position interval $[p'_1+1,p'_{s'}-1]$ to report $E$ in increasing order, skipping an empty interval.
  The total query time is $O(m+\log\log n+\out)$.
  For $m\ge\tau$, the $O(\log n\log\log n)$ query overhead is absorbed by $O(m)$.
  By Lemmas~\ref{lem:pred}, \ref{lem:sorted-segments}, and \ref{lem:shallow-nodes},
  the structures stored at nodes in $\mathcal V_\tau$ require
  $O(n\tau) \subset O(n \log^2 n)$ words and $O(n\tau\log n) \subset O(n \log^3 n)$ construction time.
\end{proof}

\section{Conclusions}
We presented indexes for reporting and counting maximal $G$-clusters,
and for reporting clusters formed by occurrences fully contained in a query substring.
The indexes use $O(n\log n)$ words for reporting and counting and $O(n\log^2 n)$ words for substring reporting.
All queries take $O(m+\log\log n)$ time, plus the output size for reporting.
A direction for future work is to reduce the space for substring reporting to $O(n\log n)$ words while preserving the query-time bound.

\begin{credits}
  \subsubsection{\ackname}
  This work was supported by JSPS KAKENHI Grant Number JP24K20734.

\end{credits}

\end{document}